\documentclass[a4paper,fleqn]{cas-dc}

\usepackage[authoryear,longnamesfirst]{natbib}
\usepackage{amsmath,amssymb,amsfonts}
\usepackage{graphicx}
\usepackage{booktabs}
\usepackage{xcolor}
\usepackage{rotating}
\usepackage{float}  
\usepackage{placeins}
\def\tsc#1{\csdef{#1}{\textsc{\lowercase{#1}}\xspace}}
\tsc{WGM}
\tsc{QE}
\begin{document}
\let\WriteBookmarks\relax
\def\floatpagepagefraction{1}
\def\textpagefraction{.001}

% Short title
\shorttitle{ViPER: Vision-based Packing-Aware Encoder}

% Short author
\shortauthors{Qaiser et al.}

% Main title
\title[mode=title]{ViPER: Vision-based Packing-Aware Encoder for Robust Malware Detection}

% First author
\author[1]{Fatima Qaiser}[orcid=0009-0005-5368-2433]
\ead{bscs2236@pieas.edu.pk}
\credit{Conceptualization, Methodology, Software, Formal analysis,
        Investigation, Data curation, Visualization,
        Writing {--} original draft}

\affiliation[1]{organization={Department of Computer \& Information Sciences,
                Pakistan Institute of Engineering \& Applied Sciences (PIEAS)},
                city={Islamabad},
                country={Pakistan}}

% Second author
\author[1]{Bisma Tahir}[orcid=0009-0008-8582-7205]
\ead{bscs2261@pieas.edu.pk}
\credit{Software, Validation, Data curation,
        Writing {--} original draft}

% Third author --- Corresponding
\author[1]{Muhammad Abid Mughal}[orcid=0000-0002-6148-9634]
\cormark[1]
\ead{mabidm@pieas.edu.pk}
\credit{Conceptualization, Resources, Supervision,
        Project administration,
        Writing {--} review and editing}

% Fourth author

% Sixth author
\author[1]{Nauman Shamim}
\ead{nauman@pieas.edu.pk}
\credit{Supervision,
        Writing {--} review and editing}

\cortext[1]{Corresponding author}

% Abstract
\begin{abstract}
Visualization-based malware detection maps raw binary bytes to grayscale
images and applies learned visual classifiers, providing an
evasion-resistant and disassembly-free alternative to conventional analysis
pipelines. However, executable packing remains a critical failure mode:
packed binaries produce high-entropy images that obscure the structural
patterns these models rely on. Because packing is also prevalent in benign
software (e.g., for compression or copy protection), packing state alone is
not a reliable indicator of maliciousness, and existing approaches do not
address this challenge within a unified supervised framework.
We present \textbf{ViPER}, a \textbf{Vi}sion-based \textbf{P}acking-Aware
\textbf{E}ncoder for \textbf{R}obust malware detection. ViPER builds on a
LoRA-adapted ViT-B/14 backbone with a dual-head architecture that jointly
learns malware classification and packing detection. A packing-aware gating
mechanism conditions malware predictions on the inferred packing state,
enabling distinct decision boundaries for packed and unpacked inputs. To
address packing label skew during training, we employ frequency-weighted
losses with stratified sampling over joint class-packing strata.
Evaluated on 200,000 Windows PE byteplot images, ViPER achieves a balanced
accuracy of 0.8521, ROC-AUC of 0.9260, and AUPR of 0.9279, outperforming
representative state-of-the-art baselines across all primary metrics, while
attaining a packing detection AUC of 0.9949.
\end{abstract}

% Highlights
\begin{highlights}
\item ViPER jointly learns malware classification and packing detection from byteplots
\item Residual gating mechanism conditions malware predictions on inferred packing state
\item LoRA-adapted ViT-B/14 trains only 1.49M of 88.66M parameters efficiently
\item Achieves ROC-AUC of 0.9260 and AUPR of 0.9279 on 200,000 PE byteplot images
\item Packing detection auxiliary head attains AUC of 0.9949 on the test set
\end{highlights}

% Keywords

\maketitle

%=============================================================================
\section{Introduction}
\label{sec:intro}
%=============================================================================

Windows malware authors have long understood that detection is fundamentally
a representation problem. If a defender cannot reliably extract features from
an executable, no classifier, however sophisticated, can compensate for
that upstream failure.Packing targets this weakness directly: when a binary's
payload is compressed or encrypted prior to delivery, static
disassembly returns little more than an unstructured byte
sequence, and the import tables, string literals, and
control-flow patterns that conventional pipelines rely on
become unavailable \citep{lyda2007using}. Dynamic analysis
can partially recover this information at runtime, yet the
computational cost per sample makes it impractical at
deployment scale, and samples that detect sandbox conditions
simply withhold their payloads until they execute in a
genuine environment.

Binary visualization, introduced by \citet{nataraj2011malware},
takes a different approach. Rather than parsing executable
structure, it assigns each byte a pixel intensity value and
arranges these values row by row into a two-dimensional grid,
producing what is commonly called a \textit{byteplot}. A
learned classifier then operates directly on this image.
Because no structural parsing is involved, the representation
remains tolerant of encryption without modification, and
malware families tend to produce consistent texture
signatures in byteplot space that persist across repackaging.

However, this framing does not account for the visual
heterogeneity that packing introduces. A packed binary,
whose payload has been compressed or encrypted, fills its
byteplot with near-uniform, high-entropy noise that carries
little structural information. An unpacked binary looks
quite different: low-entropy regions emerge wherever code
sections, import tables, or embedded strings reside in
memory. Passing both types through a single shared decision
boundary treats visually dissimilar inputs as equivalent,
and detection accuracy suffers most on the samples that have
been most aggressively obfuscated.

The problem is compounded by how widespread packing is
outside the malware ecosystem. Commercial copy-protection
tools, software installers, and archive utilities all
generate high-entropy output for entirely legitimate reasons,
which means packing state alone cannot reliably signal
malicious intent. In our dataset of 200,000 Windows PE
binaries, 76.20\% of malware samples and 73.08\% of benign
samples are packed — a near-symmetric distribution that
renders packing state alone a poor discriminator. A reliable detector
must therefore simultaneously infer whether a binary is packed
and whether it is malicious, using each inference to inform
the other rather than treating them as independent questions.

Prior work has addressed packing either by ignoring it entirely
or by running a separate unpacking tool before feature
extraction. The first approach accepts degraded performance on
obfuscated samples. The second depends on the unpacker
recognising the specific packer used, and fails silently on
custom or novel compression schemes. Neither approach integrates
packing knowledge into the learned representation itself.

Recent self-supervised vision transformers have demonstrated
that large-scale visual pre-training produces representations
that transfer across surprisingly distant
domains \citep{oquab2023dinov2}. Whether the structural gap
between natural images and byte-level visualizations is small
enough for such transfer to be useful is an open empirical
question. Parameter-efficient adaptation methods such as
LoRA \citep{hu2022lora} make it practical to fine-tune very
large backbones on domain-specific data without discarding
the pre-trained representations, keeping trainable parameter
counts manageable even when the backbone itself is large.

\subsection{Summary of Contributions}

\textbf{Dual-head packing-aware architecture.}
We propose a novel dual-head vision transformer architecture
in which packing detection serves as an auxiliary supervised
task. A learnable residual gating layer fuses packing
confidence into the malware classification pathway,
modulating the primary head logit as a function of the
inferred packing state. To the best of our knowledge, no
prior work treats packing state as an auxiliary learning
signal within a unified visual classification framework.

\textbf{Multi-task joint training.}
ViPER is trained with a combined objective in which a weighted
packing detection loss serves as an auxiliary signal alongside
the primary malware classification loss, with the auxiliary
weight set to reflect the subordinate role of packing
detection without interfering with the primary objective.

\textbf{Packing asymmetry analysis and dataset-level insight.}
We perform a dataset-level packing analysis using
Detect-It-Easy, revealing that packing prevalence is
substantial in both classes (76.20\% malware, 73.08\% benign),
indicating that packing state alone is an insufficient
discriminator and motivating the learned, texture-based
packing head design.

%=============================================================================
\section{Related Work}
\label{sec:related}
%=============================================================================

Visualization-based malware analysis, self-supervised vision
transformers, multi-task learning for security, and packing
detection form the four pillars on which ViPER is built.
This section reviews each area, highlights the gaps that
motivated our work, and ends with a consolidated comparison
in Table~\ref{tab:related}.

\subsection{Visualization-Based Malware Detection}

\citet{nataraj2011malware} made the earliest observation that
raw executable bytes, when arranged as a two-dimensional
pixel grid, produce grayscale images whose texture differs
consistently across malware families. Since pixel values come
directly from byte content rather than from any parsed
representation, the method avoids disassembly altogether and
keeps working even when the underlying binary has been
encrypted or otherwise modified.

Convolutional approaches followed shortly after.
\citet{gibert2019using} showed that CNNs learn spatial
feature hierarchies from byteplots that consistently beat
hand-crafted texture descriptors. \citet{vasan2020imcfn}
built on this with IMCFN, which fine-tunes an
ImageNet-pretrained network on the Malimg benchmark and
reaches competitive accuracy despite the large domain gap
between natural photographs and binary visualizations.
\citet{bhodia2019transfer} showed that even under tight
label budgets this gap remains manageable, as long as the
source model was pretrained at sufficient scale.

Transformer architectures entered the field as these models
became more widely available. \citet{ashawa2024enhanced}
directly compared ResNet-152 and a standard ViT on grayscale
malware images and found that accuracy on balanced benchmarks
such as Malimg paints an overly optimistic picture: the two
architectures diverge noticeably once class proportions
shift, with direct consequences for how security datasets
should be evaluated.
\citet{bavishi2024levitmc} addressed the detection and
attribution tasks jointly through LeViT-MC, a pipeline that
passes samples through a DenseNet binary detector before
handing them to a LeViT classifier for family assignment.
\citet{lu2026progressive} reported that pairing a LeViT
backbone with Progressive Focal Loss and Automatic Mixed
Precision reduces GPU memory consumption by around 25\% while
pushing Macro-F1 to 0.953 and Cohen's Kappa to 0.966 on
Malimg. Hybrid designs have also made progress: combining
ConvNeXt-Tiny with a Swin
Transformer \citep{hybrid2025convnext} produced 99.25\%
accuracy across three benchmarks, and \citet{masab2025gcvit}
found that the global context attention mechanism in GCViT is
particularly useful for separating families whose byteplot
textures are close to one another.

Despite this breadth of work, none of the methods above
account for the packing state of the binary being classified.
ViPER fills this gap by adding a packing detection head that
runs alongside the malware classifier, with a gating layer
that lets the inferred packing state influence the final
detection decision.

\subsection{Self-Supervised Vision Transformers and Transfer Learning}

The Vision Transformer (ViT), introduced by
\citet{dosovitskiy2020vit}, demonstrated that pure
attention-based architectures can match or surpass CNN
performance on image recognition benchmarks when pre-trained
at sufficient scale. The subsequent development of
self-supervised pre-training objectives, including masked
image modelling and self-distillation, has produced frozen
feature extractors that generalise across visual domains
without task-specific supervision.

\citet{caron2021dino} showed that self-distillation with no
labels produces features with emergent semantic segmentation
properties, while its successor
DINOv2 \citep{oquab2023dinov2} scaled this approach to 142
million parameters and demonstrated that frozen ViT-B/14
features achieve state-of-the-art performance on a wide range
of downstream tasks through simple linear probing. Despite the
remarkable cross-domain transfer demonstrated by these models
on natural image benchmarks, their applicability to binary
visualization has not been systematically studied. ViPER
therefore adopts a purely visual self-supervised encoder as
its backbone, motivated by the structural mismatch between
text-image supervision and the byte-level statistics of
byteplot imagery.

\subsection{Multi-Task Learning for Malware Analysis}

Multi-task learning (MTL) improves model generalisation by
sharing representations across related objectives, reducing
the risk of overfitting on any single task and providing
implicit regularisation through auxiliary
gradients \citep{caruana1997multitask}. In the context of
malware analysis, MTL has been applied to joint classification
and family attribution \citep{yan2019detecting}, where sharing
lower-layer representations between detection and attribution
heads yielded consistent gains over independently trained
models. \citet{huang2022mtnet} proposed MTNet, which jointly
optimises malware detection and API call sequence prediction,
demonstrating that auxiliary sequence modelling provides
useful inductive bias for the detection objective.
\citet{ki2015novel} incorporated packing-related metadata as
a hard pre-processing step, unpacking the binary prior to
feature extraction rather than treating it as a learned
signal, which introduces brittleness against unknown or custom
packers.

ViPER departs from both paradigms by treating packing
detection as a learned auxiliary task within a unified visual
classification framework, trained end-to-end with a joint
weighted loss. The auxiliary head learns to infer packing
state directly from byteplot visual features, and its
confidence is propagated to the primary detection head through
a residual gating mechanism.

\subsection{Packing Detection and Static Analysis Tools}

Early packing detection research rested on a simple
empirical observation: packed or encrypted executable
sections consistently show Shannon entropy above 7.0
bits \citep{lyda2007using}. This threshold is cheap to
compute and straightforward to implement, which is why
tools such as PEiD and Detect-It-Easy
\citep{horsicak2022die} adopted it as a first-pass filter,
pairing it with signature databases that match known packer
fingerprints against PE structural features.

The weaknesses of this design become apparent quickly in
practice. When \citet{ugarte2015sok} carried out a
systematic evaluation across a wide range of packing
detection tools, two recurring failure patterns emerged:
signature matching breaks down against custom or previously
unseen packers, and raw entropy thresholding generates too
many false positives because compressed archives,
self-extracting installers, and media files fall in the
same high-entropy range as packed malware without actually
being malicious. \citet{biondi2018tutorial} traced both
failures to the same underlying cause: a single scalar
entropy value throws away the spatial and structural
relationships inside a binary that would genuinely separate
packing from legitimate compression, and called for richer
feature representations capable of capturing those
relationships.

ViPER addresses this by shifting packing detection into the
visual domain. The auxiliary packing head works on the same
\texttt{[CLS]} embedding produced by the shared ViT-B/14
backbone, meaning the features used to infer packing state
are identical to those driving malware classification. This
coupling ensures that the packing signal entering the
residual gate is grounded in byteplot texture rather than
an independent entropy measurement, so the correction the
gate produces stays aligned with the primary detection
objective rather than working against it.

\subsection{Class Imbalance in Malware Detection}

Class imbalance is a pervasive challenge in cybersecurity
datasets, where benign samples typically outnumber malware
samples by large margins in real-world
deployment \citep{pendlebury2019tesseract}. Existing
approaches include oversampling via
SMOTE \citep{chawla2002smote}, cost-sensitive
learning \citep{elkan2001foundations}, and threshold
calibration at inference time. \citet{lu2026progressive}
demonstrated that Progressive Focal Loss improves
minority-class Macro-F1 and Cohen's Kappa on Malimg without
adding parameters or inference latency. ViPER addresses
packing label skew through class-frequency-weighted
cross-entropy applied independently to each head, combined
with weighted random sampling over joint class-packing strata.

\subsection{Summary and Research Gap}

Table~\ref{tab:related} consolidates representative works
across the four areas surveyed above. Two gaps are
consistently apparent: no existing work combines
visualization-based malware detection with explicit
packing-state supervision in a unified learned framework; and
self-supervised vision transformers have not been applied to
binary visualization tasks with parameter-efficient
adaptation. ViPER is designed to close both gaps
simultaneously.

%% In cas-dc, wide tables use table* to span both columns
\begin{table*}[htbp]
\caption{Comparison of Representative Related Works.}
\label{tab:related}
\begin{tabular*}{\tblwidth}{@{}p{3.2cm}llccc@{}}
\toprule
\textbf{Work} & \textbf{Input} & \textbf{Model} &
\textbf{Pack-Aware} & \textbf{Multi-Task} &
\textbf{Self-Supervised} \\
\midrule
Nataraj et al.\ \citeyearpar{nataraj2011malware}
    & Byteplot        & k-NN              & No      & No  & No \\
Gibert et al.\ \citeyearpar{gibert2019using}
    & Byteplot        & CNN               & No      & No  & No \\
Vasan et al.\ \citeyearpar{vasan2020imcfn}
    & Byteplot        & Fine-tuned CNN    & No      & No  & No \\
Ki et al.\ \citeyearpar{ki2015novel}
    & Static features & SVM               & Partial & No  & No \\
Yan et al.\ \citeyearpar{yan2019detecting}
    & Raw bytes       & Multi-task DNN    & No      & Yes & No \\
Huang et al.\ \citeyearpar{huang2022mtnet}
    & API sequences   & MTNet             & No      & Yes & No \\
Ugarte-Pedrero et al.\ \citeyearpar{ugarte2015sok}
    & PE metadata     & Entropy threshold & Yes     & No  & No \\
Ashawa et al.\ \citeyearpar{ashawa2024enhanced}
    & Byteplot        & ResNet-152 + ViT  & No      & No  & No \\
Bavishi \& Narayanan\ \citeyearpar{bavishi2024levitmc}
    & Byteplot        & LeViT-MC          & No      & No  & No \\
Alshomrani et al.\ \citeyearpar{hybrid2025convnext}
    & Byteplot        & ConvNeXt + Swin   & No      & No  & No \\
Masab et al.\ \citeyearpar{masab2025gcvit}
    & Byteplot        & GCViT             & No      & No  & No \\
Lu et al.\ \citeyearpar{lu2026progressive}
    & Byteplot        & LeViT+Focal Loss  & No      & No  & No \\
\midrule
\textbf{ViPER (ours)}
    & \textbf{Byteplot} & \textbf{ViT-B/14 + LoRA}
    & \textbf{Learned} & \textbf{Yes} & \textbf{Yes} \\
\bottomrule
\end{tabular*}
\end{table*}

%=============================================================================
\section{Proposed Methodology}
\label{sec:method}
%=============================================================================

\begin{sidewaysfigure*}
    \centering
    \includegraphics[width=0.85\textheight]{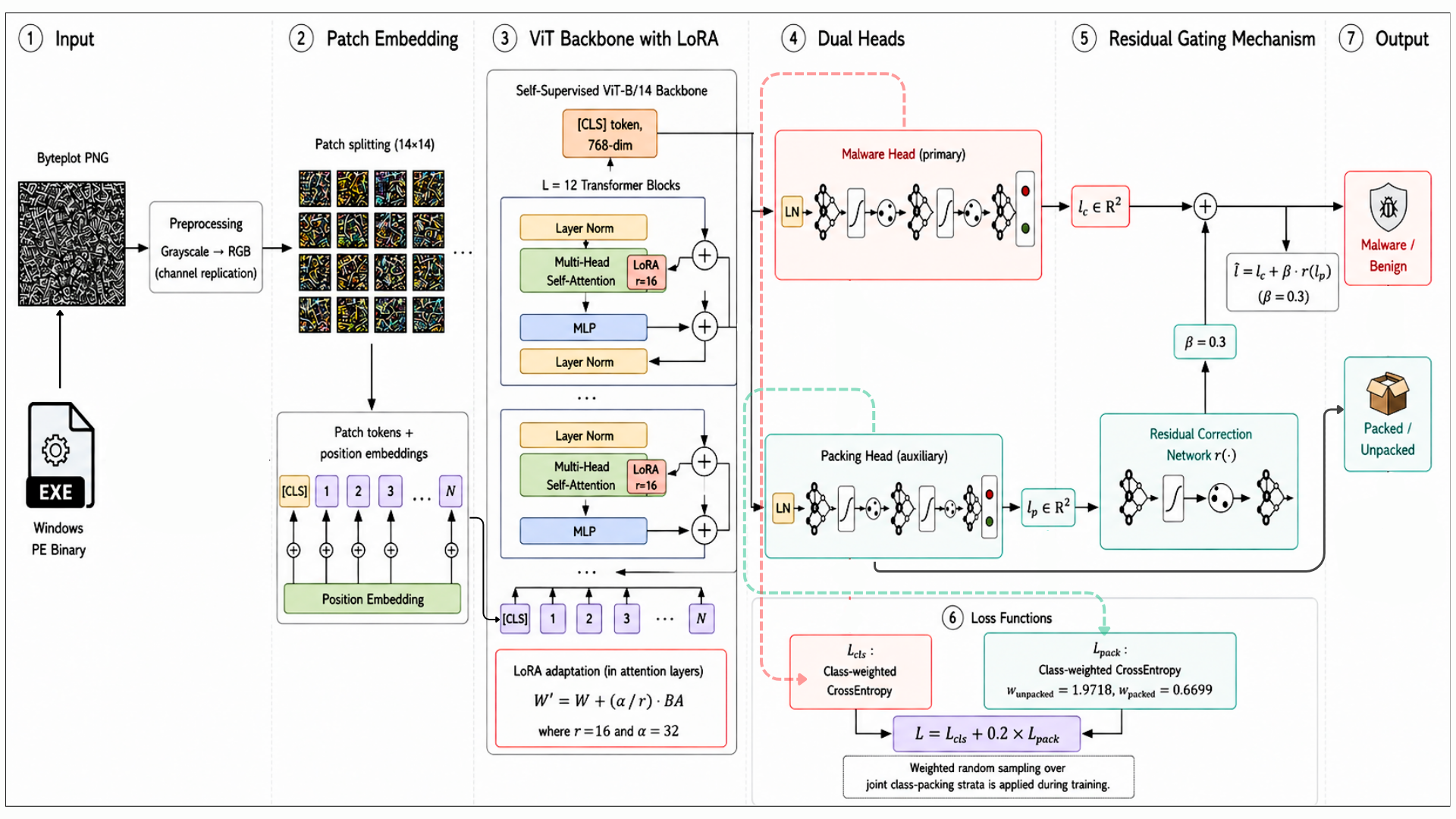}
    \caption{Overall architecture of ViPER. A LoRA-adapted
    DINOv2 ViT-B/14 backbone produces a shared \texttt{[CLS]}
    embedding fed to two independent MLP heads. The packing
    head output passes through a residual gating network whose
    output is added as a weighted correction to the malware
    classification logit. Both heads are trained jointly with
    class-weighted cross-entropy losses.}
    \label{fig:architecture}
\end{sidewaysfigure*}

We propose \textbf{ViPER} (\textbf{Vi}sion-based
\textbf{P}acking-Aware \textbf{E}ncoder for \textbf{R}obust
Malware Detection), a dual-head vision transformer framework
that jointly learns malware classification and packing
detection from byteplot visualizations of Windows PE binaries.
The overall architecture is illustrated in
Fig.~\ref{fig:architecture}.

%% sidewaysfigure* spans both columns in cas-dc

\subsection{Problem Formulation}

Let $\mathcal{D} = \{(\mathbf{x}_i, y_i^c,
y_i^p)\}_{i=1}^{N}$ denote the dataset of $N$ byteplot
images, where
$\mathbf{x}_i \in \mathbb{R}^{3 \times 224 \times 224}$
is the $i$-th input image, $y_i^c \in \{0, 1\}$ is the
malware/benign label ($0$ = benign, $1$ = malware), and
$y_i^p \in \{0, 1\}$ is the packing label ($0$ = unpacked,
$1$ = packed).

ViPER learns a shared visual encoder
$f_\theta : \mathbb{R}^{3 \times 224 \times 224} \to
\mathbb{R}^{768}$ together with two classification heads and
a residual gating mechanism:
\begin{align}
    \mathbf{z}       &= f_\theta(\mathbf{x})
    \label{eq:encoder} \\
    \mathbf{l}_c     &= g_c(\mathbf{z})
    \label{eq:clshead} \\
    \mathbf{l}_p     &= g_p(\mathbf{z})
    \label{eq:packhead} \\
    \hat{\mathbf{l}} &= \mathbf{l}_c +
                        \beta \cdot r(\mathbf{l}_p)
    \label{eq:gated}
\end{align}
where $g_c$ and $g_p$ are the malware and packing MLP heads
respectively, $r(\cdot)$ is a residual correction network
mapping packing logits to a correction in the same output
space as $\mathbf{l}_c$, and $\beta = 0.3$ is a fixed scalar
weight. The final prediction is
$\hat{y}^c = \arg\max(\hat{\mathbf{l}})$.

The model is trained to minimise the multi-task weighted loss:
\begin{equation}
    \mathcal{L} =
    \mathcal{L}_{\text{cls}}(\hat{\mathbf{l}},\, y^c)
    + \lambda\,
    \mathcal{L}_{\text{pack}}(\mathbf{l}_p,\, y^p)
    \label{eq:loss}
\end{equation}
where $\mathcal{L}_{\text{cls}}$ and
$\mathcal{L}_{\text{pack}}$ are class-weighted cross-entropy
losses and $\lambda = 0.2$ is the auxiliary loss weight.

\subsection{Dataset}

\subsubsection{Source and Composition}
The malware corpus was assembled from publicly available
malware repositories. Due to the sensitive nature of
executable samples, the raw binaries are not publicly
released. To facilitate reproducibility, the SHA-256 hashes
of all 200,000 Windows PE binaries used in this study are
publicly available\footnote{\url{https://drive.google.com/drive/folders/1uzVXm7ceyjR0Gae6ArT4kGotcB-5uVHM?usp=sharing}},
allowing verification and reconstruction of the dataset from
publicly available malware repositories such as VirusTotal
or MalwareBazaar. The dataset comprises 200,000 PNG
images: 100,000 malware samples and 100,000 benign samples,
covering a diverse range of Windows Portable Executable (PE)
binaries. Each image is a grayscale byteplot in which raw
binary bytes are mapped to pixel intensity values and arranged
into a two-dimensional grid row by
row \citep{nataraj2011malware}. This representation requires
neither disassembly nor dynamic execution, making it
inherently robust to many evasion strategies.

\subsubsection{Packing Label Generation}
Ground-truth packing labels are not provided with the dataset.
We derive them automatically using Detect-It-Easy
(DiE) \citep{horsicak2022die}, an open-source static analysis
tool that combines signature-based packer identification with
heuristic structural analysis of PE files. For each binary,
DiE examines the PE section layout, section name conventions,
entry-point characteristics, and overlay patterns, and cross
references these structural indicators against an extensible
database of known packer signatures to produce a packing
verdict. Samples for which DiE reports a known packer,
protector, or installer signature are labelled
\textit{packed}; samples for which no such indicator is
detected are labelled \textit{unpacked}. This approach grounds
the packing labels in interpretable structural evidence rather
than scalar entropy values, which are known to produce
excessive false positives due to legitimately high-entropy
benign content such as compressed archives and media
files \citep{ugarte2015sok,biondi2018tutorial}.

\begin{figure*}[htbp]
    \centering
    \includegraphics[width=0.32\textwidth]{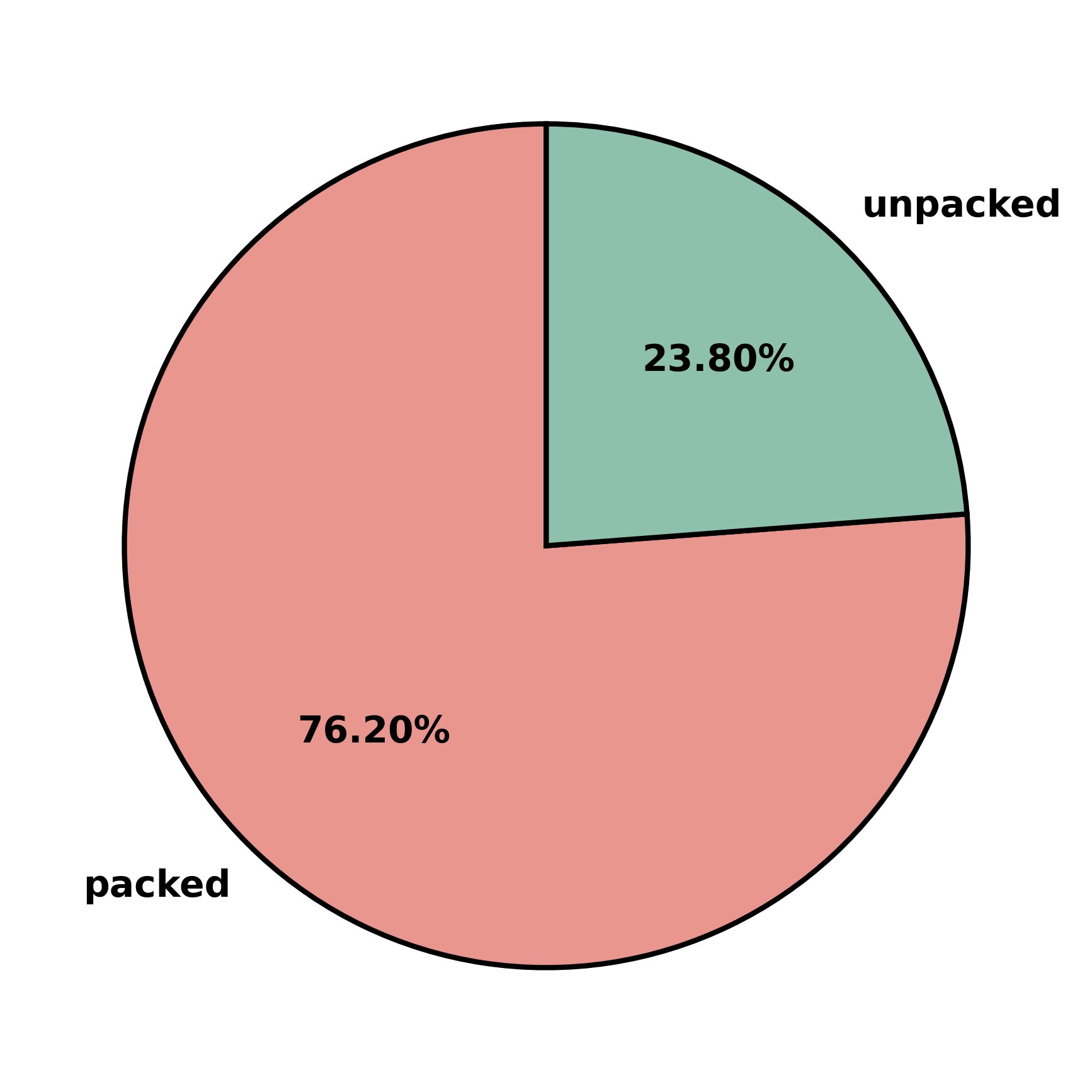}
    \hspace{1cm}
    \includegraphics[width=0.32\textwidth]{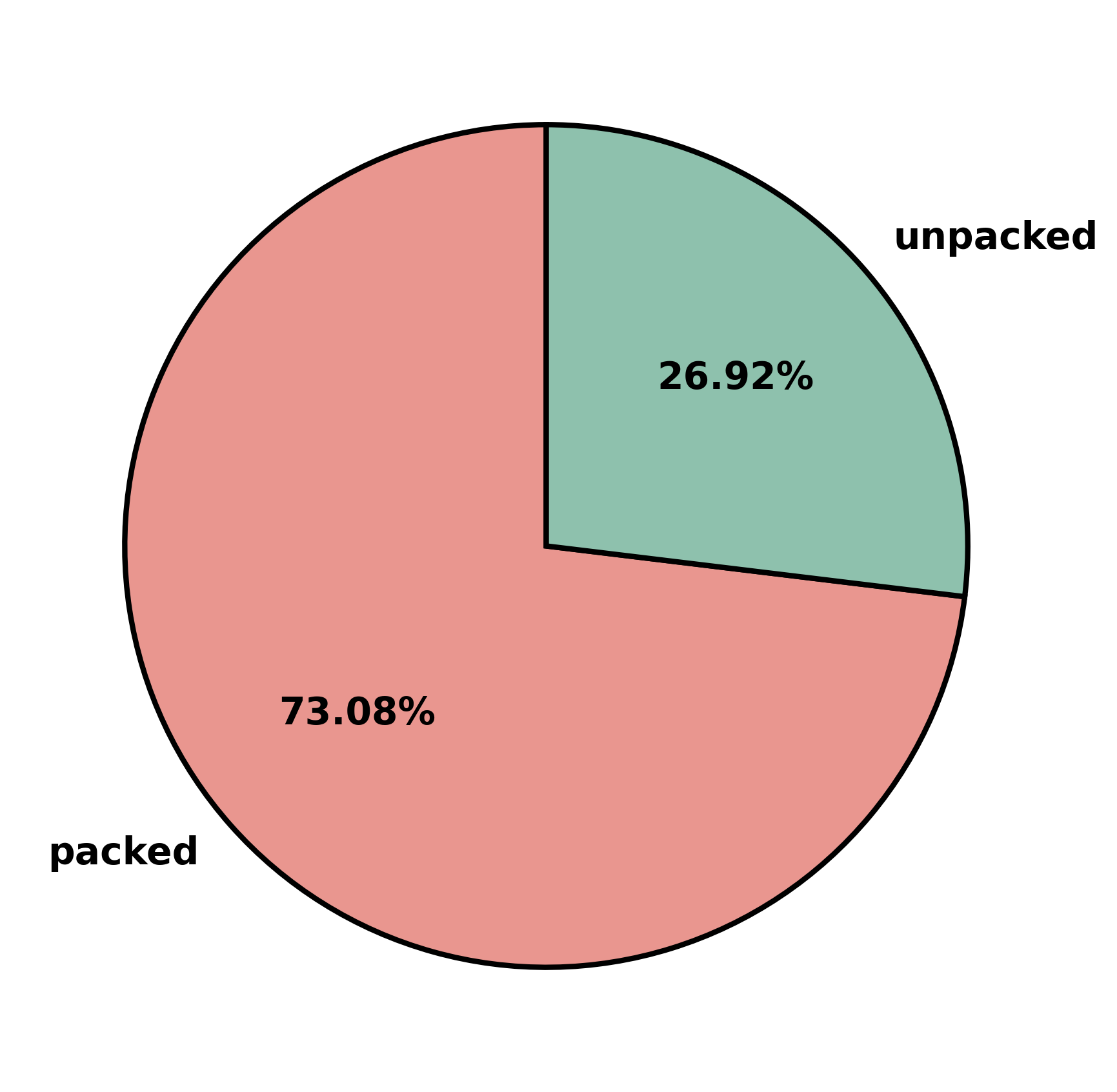}
    \caption{Packing label distribution in the dataset.
    \textbf{Left:} Malware samples (76.20\% packed).
    \textbf{Right:} Benign samples (73.08\% packed). The
    near-symmetric packing prevalence across both classes
    confirms that packing state alone is an insufficient
    discriminator.}
    \label{fig:pie_charts}
\end{figure*}

The resulting packing distribution is summarised in
Table~\ref{tab:packing_dist} and visualised in
Fig.~\ref{fig:pie_charts}. Among malware samples, 76.20\%
are packed; among benign samples, 73.08\% are packed. The
near-symmetric packing prevalence across both classes
indicates that packing state alone is an insufficient
discriminator, and motivates the residual gating design: the
gate provides a soft texture-based correction rather than a
hard routing signal.

%% figure* spans both columns in cas-dc

%% Single-column table — use plain {table} in cas-dc
\begin{table}[htbp]
\caption{Packing Label Distribution in the Malware Dataset}
\label{tab:packing_dist}
\centering
\begin{tabular}{lrrrr}
\toprule
\textbf{Class} & \textbf{Total} & \textbf{Packed} &
\textbf{Unpacked} & \textbf{Pack \%} \\
\midrule
Benign  & 100,000 & 73,085  & 26,915 & 73.08\% \\
Malware & 100,000 & 76,200  & 23,800 & 76.20\% \\
\midrule
Overall & 200,000 & 149,285 & 50,715 & 74.64\% \\
\bottomrule
\end{tabular}
\end{table}

\subsubsection{Dataset Partitioning}
The dataset is partitioned into training (70\%), validation
(15\%), and test (15\%) sets using a two-stage stratified
split. A composite stratum label is constructed by
concatenating the binary class label and the packing label,
yielding four strata: \textit{benign-unpacked},
\textit{benign-packed}, \textit{malware-unpacked}, and
\textit{malware-packed}. \texttt{StratifiedShuffleSplit} from
scikit-learn is applied twice to ensure that all four strata
are proportionally represented in every partition, preventing
the inadvertent introduction of evaluation bias. The resulting
partition sizes are summarised in Table~\ref{tab:splits}.

\begin{table}[htbp]
\caption{Dataset Partition Statistics}
\label{tab:splits}
\centering
\begin{tabular}{lrrrr}
\toprule
\textbf{Split} & \textbf{Total} & \textbf{Benign} &
\textbf{Malware} & \textbf{Packed} \\
\midrule
Train & 140,000 & 70,000  & 70,000  & 104,500 \\
Val   &  30,000 & 15,000  & 15,000  &  22,393 \\
Test  &  30,000 & 15,000  & 15,000  &  22,392 \\
\midrule
Total & 200,000 & 100,000 & 100,000 & 149,285 \\
\bottomrule
\end{tabular}
\end{table}

%-------------------------------------------------------------
\subsection{Data Preprocessing}

\subsubsection{Image Preparation}
Byteplot images are stored as grayscale PNGs, as each pixel
encodes a single byte value in the range $[0, 255]$ and no
colour information is present in the source data. To satisfy
the three-channel input requirement of the ViT-B/14 backbone,
each grayscale image is converted to RGB via channel
replication, in which the single luminance channel is copied
identically across all three channels. This operation is
lossless and introduces no spurious colour information; it is
a standard adaptation step when applying pre-trained RGB
vision models to single-channel binary visualizations. All
images are resized to $224 \times 224$ pixels using bilinear
interpolation and normalised with byteplot-specific statistics
($\mu = [0.5, 0.5, 0.5]$, $\sigma = [0.25, 0.25, 0.25]$),
which are appropriate for the approximately uniform byte
distribution of binary data and preferred over ImageNet
statistics for this domain.

\subsubsection{Training Augmentation}
During training, random horizontal flip ($p = 0.5$), random
vertical flip ($p = 0.5$), colour jitter with brightness and
contrast perturbation of $\pm 0.1$, and random erasing
($p = 0.15$, scale $(0.02, 0.06)$) are applied. Spatial
flips are valid augmentations for byteplots because the
mapping from byte sequence to pixel coordinates depends only
on the chosen row width and is otherwise spatially arbitrary.
Random rotation is deliberately excluded, as byteplots are not
rotationally invariant. No augmentation is applied during
validation or testing.

%-------------------------------------------------------------
\subsection{Class Imbalance Handling}

The packing label distribution in the training set is
imbalanced (74.64\% packed, 25.36\% unpacked). To address
this, the packing head loss is independently weighted using
balanced class weights computed from the training partition:
\begin{equation}
    w_p =
    \frac{N_{\text{train}}}{2 \cdot N_p^{\text{train}}},
    \quad p \in \{\text{unpacked},\, \text{packed}\}
    \label{eq:packweight}
\end{equation}
yielding $w_{\text{unpacked}} = 1.9718$ and
$w_{\text{packed}} = 0.6699$. The malware classification head
uses the same class-frequency weighting formulation, ensuring
the framework generalises directly to deployment settings with
realistic benign-to-malware skew without modification.

\subsubsection{Weighted Random Sampling}
To ensure balanced batch composition during training, a
\texttt{WeightedRandomSampler} assigns each training sample a
weight inversely proportional to the frequency of its
composite stratum (class $\times$ packing label). Sampling is
performed with replacement, ensuring that all four strata are
represented in every training batch. This data-level
correction operates independently of and complementarily to
the loss-level weighting described above.

%-------------------------------------------------------------
\subsection{ViPER Architecture}
\label{sec:architecture}

\subsubsection{Backbone: Self-Supervised Visual Encoder}
ViPER employs a self-supervised vision transformer
pre-trained on 142 million images via self-distillation with
no labels \citep{oquab2023dinov2} as its visual backbone. The
ViT-B/14 variant uses a patch size of $14 \times 14$ pixels
and produces a 768-dimensional \texttt{[CLS]} token embedding
$\mathbf{z} \in \mathbb{R}^{768}$ from each input image. The
\texttt{[CLS]} token serves as the global image
representation and is fed to both classification heads.

\subsubsection{Parameter-Efficient Fine-Tuning via LoRA}
To adapt the backbone to the byteplot domain while preserving
pre-trained representations and minimising trainable
parameters, we apply Low-Rank Adaptation
(LoRA) \citep{hu2022lora} to the query-key-value (QKV)
projection matrices of all 12 transformer blocks. For each
frozen weight matrix $\mathbf{W} \in \mathbb{R}^{d \times k}$,
LoRA introduces two low-rank matrices
$\mathbf{A} \in \mathbb{R}^{r \times k}$ and
$\mathbf{B} \in \mathbb{R}^{d \times r}$ such that the
adapted forward pass computes:
\begin{equation}
    \mathbf{W}'\mathbf{x} =
    \mathbf{W}\mathbf{x}
    + \frac{\alpha}{r}\,
    \mathbf{B}\mathbf{A}\mathbf{x}
    \label{eq:lora}
\end{equation}
where $r = 16$ is the rank, $\alpha = 32$ is the scaling
factor, and all elements of $\mathbf{B}$ are initialised to
zero, ensuring an identity mapping at initialisation. All
original backbone weights remain frozen throughout training;
only $\mathbf{A}$ and $\mathbf{B}$ are updated. This
configuration yields a substantially reduced set of trainable
backbone parameters relative to full fine-tuning, enabling
efficient domain adaptation without modifying the pre-trained
feature representations.

\subsubsection{Dual Classification Heads}
Two independent MLP heads are attached to the shared backbone
embedding $\mathbf{z}$:

\textbf{Malware head} (primary task):
\begin{equation}
    g_c(\mathbf{z}) =
    \mathbf{W}_3\,\text{GELU}(
    \mathbf{W}_2\,\text{GELU}(
    \mathbf{W}_1\,\text{LN}(\mathbf{z})
    ))
    \label{eq:clshead_eq}
\end{equation}
with dimensions $768 \to 512 \to 256 \to 2$ and dropout
rates of 0.3 and 0.15 after the first and second projections
respectively.

\textbf{Packing head} (auxiliary task):
\begin{equation}
    g_p(\mathbf{z}) =
    \mathbf{W}_6\,\text{GELU}(
    \mathbf{W}_5\,\text{GELU}(
    \mathbf{W}_4\,\text{LN}(\mathbf{z})
    ))
    \label{eq:packhead_eq}
\end{equation}
with dimensions $768 \to 384 \to 192 \to 2$ and dropout rate
of 0.3. LN denotes Layer Normalisation.

\subsubsection{Packing-Aware Residual Gating Mechanism}
The packing head logit $\mathbf{l}_p \in \mathbb{R}^2$ is
passed through a residual correction network $r(\cdot)$
(Linear$(2,32) \to$ ReLU $\to$ Dropout$(0.1) \to$
Linear$(32,2)$) to produce a correction vector in the same
output space as the malware logit. The corrected logit is
computed as:
\begin{equation}
    \hat{\mathbf{l}} = \mathbf{l}_c +
    \beta \cdot r(\mathbf{l}_p), \quad \beta = 0.3
    \label{eq:residual_gate}
\end{equation}
This residual formulation ensures that the packing signal
augments rather than replaces the malware classification,
maintaining the primacy of the visual features learned by the
malware head. The choice of additive rather than
multiplicative gating is motivated by the similar packing
distributions across classes in this dataset: a multiplicative
gate conditioned on packing confidence would produce noisy
scaling and degrade the primary objective under near-symmetric
packing prevalence.

%-------------------------------------------------------------
\subsection{Training Protocol}

\subsubsection{Loss Function}
ViPER is trained with the multi-task objective defined in
Eq.~(\ref{eq:loss}).
The auxiliary weight $\lambda = 0.2$ reflects the subordinate role of packing
detection relative to the primary malware classification objective.
This value is set conservatively to account for the observation that packing
distributions are similar across classes in this dataset, reducing the
discriminative value of the auxiliary signal and warranting a lower weight
to prevent interference with the primary head.

\subsubsection{Optimiser and Schedule}
Parameters are updated using AdamW \citep{loshchilov2019adamw} with
differential learning rates: LoRA adapter parameters are trained at
$\eta_{\text{LoRA}} = 1 \times 10^{-4}$ and classification head parameters
at $\eta_{\text{head}} = 3 \times 10^{-4}$, both with weight decay $10^{-2}$.
The learning rate follows a linear warmup over three epochs followed by
cosine annealing \citep{loshchilov2017sgdr} over the remaining training
budget, with a minimum learning rate of $\eta_{\min} = 10^{-7}$.
Gradient norms are clipped to 1.0 to stabilise transformer fine-tuning.

\subsubsection{Training Configuration}
Training proceeds for a maximum of 20 epochs with early stopping triggered
when the validation ROC-AUC of the malware head does not improve for five
consecutive epochs.
The model checkpoint achieving the highest validation AUC is retained for all
test evaluations.
A per-GPU batch size of 128 with BFloat16 mixed-precision training is used
throughout.
All experiments are conducted on an NVIDIA RTX 5090 GPU (32\,GB VRAM).
Table~\ref{tab:hyperparams} summarises all hyperparameters.

\begin{table}[htbp]
\caption{ViPER Hyperparameter Summary}
\label{tab:hyperparams}
\begin{tabular*}{\tblwidth}{@{}LL@{}}
\toprule
\textbf{Hyperparameter} & \textbf{Value} \\
\midrule
Backbone              & ViT-B/14 (DINOv2 self-supervised) \\
LoRA rank $r$         & 16 \\
LoRA scaling $\alpha$ & 32 \\
Input resolution      & $224 \times 224$ \\
Normalisation         & mean 0.5, std 0.25 (byteplot) \\
Batch size            & 128 \\
Max epochs            & 20 \\
Best epoch            & 18 \\
Early stopping        & patience = 5 (val AUC) \\
Optimiser             & AdamW \\
LR (LoRA)             & $1 \times 10^{-4}$ \\
LR (heads)            & $3 \times 10^{-4}$ \\
Weight decay          & $10^{-2}$ \\
LR warmup             & 3 epochs (linear) \\
LR schedule           & Cosine annealing \\
Grad clip             & 1.0 \\
$\lambda$ (pack)      & 0.2 \\
$\beta$ (gate)        & 0.3 \\
Dropout (cls)         & 0.3 / 0.15 \\
Dropout (pack)        & 0.3 \\
Precision             & BFloat16 \\
Hardware              & NVIDIA RTX 5090 (32\,GB) \\
\bottomrule
\end{tabular*}
\end{table}

%=============================================================================
\section{Results and Discussion}
\label{sec:results}
%=============================================================================

\subsection{Evaluation Protocol}

All models are evaluated on the same held-out test set of 30,000 samples
(15,000 benign, 15,000 malware) derived from the joint-stratified split
described in Section~\ref{sec:method}.
We report balanced accuracy (BalAcc), ROC-AUC (AUC), and area under the
precision-recall curve (AUPR).
Balanced accuracy and AUPR are prioritised over plain accuracy and AUC
respectively, as they are more informative under class imbalance conditions.
All metrics are computed on the test set using the best validation-AUC
checkpoint.

%-------------------------------------------------------------
\subsection{Ablation Study}

To isolate the contribution of each architectural component, we conduct a
controlled ablation across four configurations with all other settings held
constant.
Table~\ref{tab:ablation} reports test set results for each variant.

\begin{table*}[htbp]
\caption{Ablation Study on the Malware Dataset Test Set (30,000 samples).
Bal.\ Acc.\ = balanced accuracy; AUPR = area under precision-recall curve.}
\label{tab:ablation}
\centering
\begin{tabular*}{\tblwidth}{@{\extracolsep{\fill}} lrrr @{}}
\toprule
\textbf{Configuration} & \textbf{BalAcc} & \textbf{AUC} & \textbf{AUPR} \\
\midrule
(1) Frozen ViT, single head          & 0.8133 & 0.8959 & 0.9049 \\
(2) LoRA, single head (no packing)   & 0.8496 & 0.9244 & 0.9273 \\
(3) LoRA, dual head (no gate)        & 0.8521 & 0.9243 & 0.9249 \\
(4) ViPER (LoRA + dual head + gate)  & \textbf{0.8521} & \textbf{0.9260} & \textbf{0.9279} \\
\bottomrule
\end{tabular*}
\end{table*}

\begin{figure*}[htbp]
    \centering
    \includegraphics[width=0.85\columnwidth]{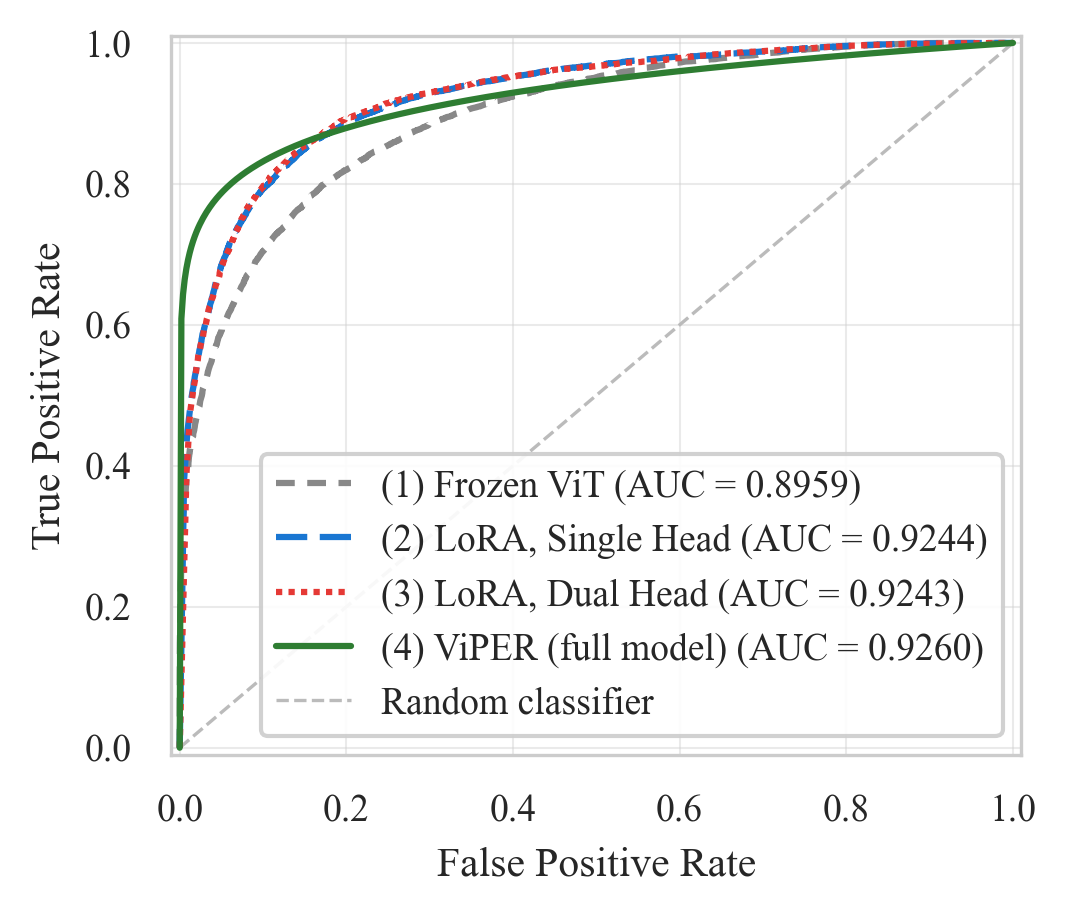}
    \caption{ROC curves for all four ablation configurations on the
    test set. The area under each curve corresponds to the AUC values
    reported in Table~\ref{tab:ablation}.}
    \label{fig:roc_overlay}
\end{figure*}

\begin{figure*}[htbp]
    \centering
    \includegraphics[width=0.75\textwidth]{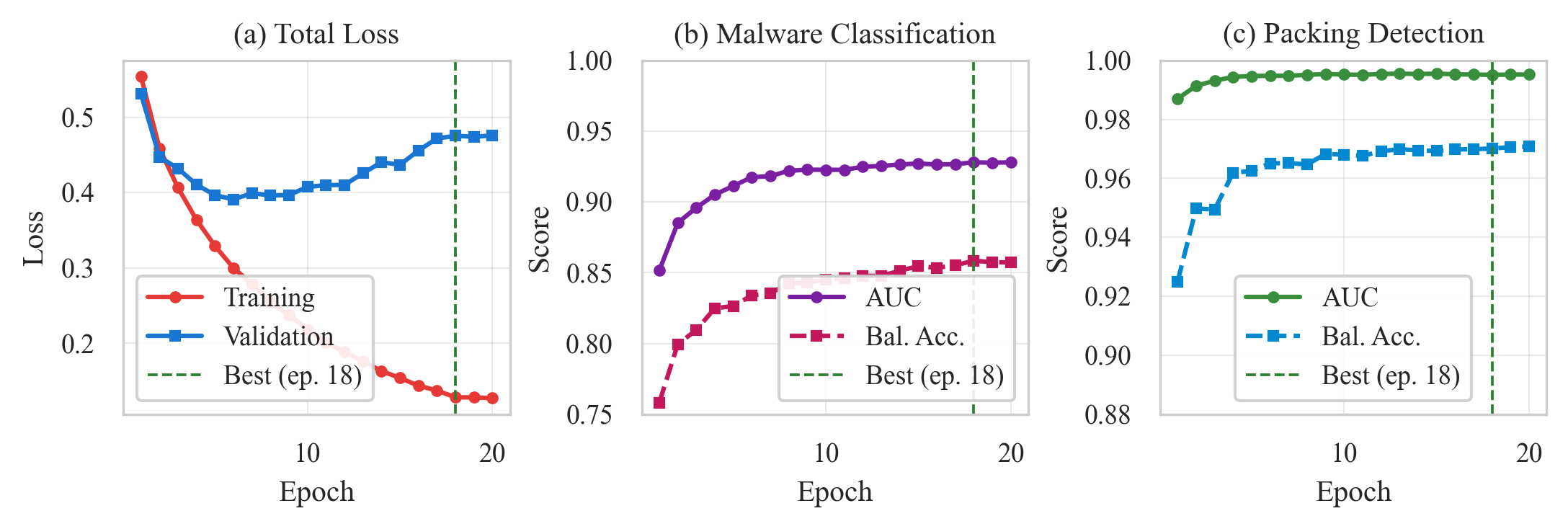}
    \caption{Training curves for ViPER (configuration 4). Loss decreases
    smoothly for both heads across 20 epochs. Validation AUC and balanced
    accuracy improve consistently through epoch 18 (best checkpoint) before
    plateauing.}
    \label{fig:training_curves}
\end{figure*}

Configuration (1) establishes the performance ceiling of frozen DINOv2
features without any fine-tuning or task-specific heads, quantifying the
baseline transfer capability of the pre-trained backbone.
Configuration (2) introduces LoRA fine-tuning and the primary malware
classification head without packing supervision, isolating the gain from
domain adaptation alone.
The jump from (1) to (2), a gain of $+$0.0363 in BalAcc and $+$0.0285 in AUC,
is the largest single gain across the ablation, confirming that LoRA
fine-tuning of the attention projections is the most impactful component.
Configuration (3) adds the dual-head multi-task objective but removes the
residual gate, allowing assessment of whether packing-aware gradient sharing
provides benefit independent of gating.
The full ViPER model (4) combines all components, achieving the best AUC
(0.9260) and AUPR (0.9279), confirming the incremental contribution of the
residual gate.
Notably, the gate improves AUC by $+$0.0017 over dual-head without gating
while maintaining identical BalAcc, suggesting the gate's benefit manifests
primarily in ranking quality rather than threshold-dependent accuracy.
The ROC curves for all four configurations are shown in
Fig.~\ref{fig:roc_overlay}, and the training dynamics of
the full model are plotted in Fig.~\ref{fig:training_curves}.

%-------------------------------------------------------------
\subsection{Main Results}

Table~\ref{tab:main_results} presents the final evaluation of the full ViPER
model on the test set for both the malware detection and packing detection
tasks, alongside the frozen baseline.

\begin{table}[htbp]
\caption{ViPER Final Test Set Results on the Malware Dataset Test Set (30,000 samples).}
\label{tab:main_results}
\centering
\begin{tabular}{lrrr}
\toprule
\textbf{Task / Model} & \textbf{BalAcc} & \textbf{AUC} & \textbf{AUPR} \\
\midrule
\multicolumn{4}{l}{\textit{Malware detection}} \\
\quad Frozen ViT (baseline)  & 0.8133 & 0.8959 & 0.9049 \\
\quad ViPER (proposed)       & \textbf{0.8521} & \textbf{0.9260} & \textbf{0.9279} \\
\quad $\Delta$ improvement   & $+$0.0388 & $+$0.0301 & $+$0.0230 \\
\midrule
\multicolumn{4}{l}{\textit{Packing detection (ViPER auxiliary head)}} \\
\quad ViPER pack head        & 0.9688 & 0.9949 & -- \\
\bottomrule
\end{tabular}
\end{table}

Table~\ref{tab:per_class} reports per-class precision, recall, and F1 for the
malware detection task.

\begin{table}[htbp]
\caption{Per-Class Malware Detection Results on the Malware Dataset Test Set
(ViPER, 30,000 samples).}
\label{tab:per_class}
\centering
\begin{tabular}{lrrrr}
\toprule
\textbf{Class} & \textbf{Precision} & \textbf{Recall} & \textbf{F1} & \textbf{Support} \\
\midrule
Benign    & 0.8426 & 0.8661 & 0.8542 & 15,000 \\
Malware   & 0.8622 & 0.8382 & 0.8500 & 15,000 \\
\midrule
Macro avg & 0.8524 & 0.8521 & 0.8521 & 30,000 \\
\bottomrule
\end{tabular}
\end{table}

%-------------------------------------------------------------
\subsection{Comparison with CNN Baselines}

To contextualise ViPER's performance within the broader landscape of
convolutional architectures, we evaluate two standard CNN baselines on the
same dataset, training protocol, and evaluation conditions:
ResNet-50 \citep{he2016resnet} with its final \texttt{layer4} block
fine-tuned, and MobileNetV3-Large \citep{howard2019mobilenetv3} with its
full backbone fine-tuned.
Both baselines use the identical AdamW optimiser with warmup and cosine
annealing schedule, the same 20-epoch training budget, and the same
class-weighted cross-entropy loss as ViPER, ensuring a controlled
comparison.

\begin{table}[htbp]
\caption{Comparison of ViPER Against CNN Baselines on the Test Set
(30,000 samples). Trainable parameters for ViPER reflect LoRA adapters
and classification heads only.}
\label{tab:cnn_comparison}
\resizebox{\columnwidth}{!}{%
\begin{tabular}{@{}lrrc@{}}
\toprule
\textbf{Model} & \textbf{Total Params} & \textbf{Trainable} & \textbf{BalAcc} \\
\midrule
MobileNetV3-Large     & 3.60M   & 3.19M (88.7\%)  & 0.8211 \\
ResNet-50             & 24.69M  & 16.15M (65.4\%) & 0.8249 \\
\midrule
\textbf{ViPER (ours)} & \textbf{88.66M} & \textbf{1.49M (1.69\%)} & \textbf{0.8521} \\
\midrule
$\Delta$ vs.\ ResNet-50   & -- & -- & $+$0.0272 \\
$\Delta$ vs.\ MobileNetV3 & -- & -- & $+$0.0310 \\
\bottomrule
\end{tabular}%
}

\vspace{4pt}

\resizebox{\columnwidth}{!}{%
\begin{tabular}{@{}lccc@{}}
\toprule
\textbf{Model} & \textbf{AUC} & \textbf{AUPR} & \textbf{TPR@1\%FPR} \\
\midrule
MobileNetV3-Large     & 0.9004 & 0.9080 & 0.4010 \\
ResNet-50             & 0.9021 & 0.9094 & 0.4201 \\
\midrule
\textbf{ViPER (ours)} & \textbf{0.9260} & \textbf{0.9279} & \textbf{0.4635} \\
\midrule
$\Delta$ vs.\ ResNet-50   & $+$0.0239 & $+$0.0185 & $+$0.0434 \\
$\Delta$ vs.\ MobileNetV3 & $+$0.0256 & $+$0.0199 & $+$0.0625 \\
\bottomrule
\end{tabular}%
}
\end{table}

Several observations merit discussion.
First, ViPER substantially outperforms both CNN baselines across all
reported metrics despite training only 1.49M parameters, compared to
16.15M for ResNet-50 and 3.19M for MobileNetV3-Large.
This result demonstrates that parameter-efficient adaptation of a
high-capacity self-supervised backbone is a more effective strategy for
this task than extensive fine-tuning of a smaller purpose-built CNN.

Second, ResNet-50 and MobileNetV3-Large achieve broadly comparable
performance (AUC of 0.9021 and 0.9004 respectively), suggesting that
architectural scale within the CNN family provides diminishing returns
at this dataset size once the backbone receptive field is sufficiently
large to capture byteplot texture statistics.
This plateau motivates the shift to transformer-based architectures
with global attention, which ViPER exploits through the DINOv2 backbone.

Third, the TPR@1\%FPR gap is operationally significant: ViPER achieves
a true positive rate of 0.4635 at a 1\% false positive rate, compared to
0.4201 for ResNet-50 and 0.4010 for MobileNetV3-Large.
In a deployment scenario where the false positive budget is constrained,
ViPER detects approximately 4.3 to 6.3 percentage points more malware
samples per 100 queried files than either CNN baseline, a meaningful
operational improvement at no additional inference cost relative to
ResNet-50 given the compact trainable footprint.

Fourth, and most relevant to the novelty of this work, the CNN baselines
have no mechanism for packing-aware inference: they apply a single shared
decision boundary regardless of the estimated packing state of the input binary.
ViPER's residual gating mechanism provides a soft, learned correction
conditioned on packing texture, which contributes to the consistent AUC
and AUPR improvements visible in Table~\ref{tab:cnn_comparison}.

%-------------------------------------------------------------
\subsection{Discussion}

The ablation results confirm that LoRA fine-tuning provides the largest single
performance gain over the frozen baseline, as domain adaptation of the
attention weights allows the model to develop byteplot-specific feature
selectivity unavailable in the natural-image pre-trained representations.
The packing-aware multi-task objective provides a secondary improvement,
particularly in AUC and AUPR, which are most sensitive to the quality of
probability calibration across the full operating range.
The residual gate further refines this, adding a directed correction from
packing-specific visual features without displacing the primary malware head.

A key observation from the dataset analysis is that packing is prevalent
in both classes (76.20\% malware, 73.08\% benign), meaning entropy alone
cannot discriminate malware from benign files.
However, the packing \textit{texture patterns} within each class remain
visually distinct: packed malware tends toward uniform high-entropy noise
associated with encryption and custom compression, while packed benign
software, such as installers and archives, exhibits more structured
high-entropy patterns.
The learned packing head is therefore forced to model fine-grained texture
differences beyond raw entropy, and the resulting correction signal provides
useful refinement to the malware head without requiring explicit supervision
on packer types.

The joint stratification of the dataset split over class-packing
strata ensures that test set performance reflects the true
operational distribution of all four subgroups, preventing
systematic evaluation bias at both the class and packing
boundaries.

%-------------------------------------------------------------
\subsection{Hyperparameter Sensitivity Analysis}

To validate the robustness of the two fixed scalar hyperparameters,
$\beta$ (residual gate weight) and $\lambda$ (auxiliary packing loss
weight), we conduct a coarse-grid sensitivity analysis. The sensitivity
runs are conducted as independent training experiments with different
random seeds from the main evaluation; the modest deviation from the
main results in Tables~\ref{tab:ablation} and~\ref{tab:main_results} reflects
stochastic training variance across runs, consistent with the narrow
stability range observed across the $\beta$ sweep itself.
The gate weight $\beta$ is varied across five values while holding all
other parameters fixed and re-evaluating the trained ViPER checkpoint;
no retraining is required since $\beta$ appears only in the forward
pass.
The auxiliary weight $\lambda$ is evaluated via independent
five-epoch training runs from scratch for each candidate value,
providing a relative comparison of training dynamics under different
auxiliary loss scales.
Results are reported in Table~\ref{tab:sensitivity}.

\begin{table}[htbp]
\caption{Sensitivity Analysis of Gate Weight $\beta$ and Auxiliary
Loss Weight $\lambda$. Metrics are reported on the held-out test set.
The trained values used in ViPER are indicated with $(\star)$.}
\label{tab:sensitivity}
\centering
\begin{tabular}{rrrr}
\toprule
$\beta$ & \textbf{BalAcc} & \textbf{AUC} & \textbf{AUPR} \\
\midrule
0.1           & 0.8567 & 0.9275 & 0.9297 \\
0.2           & 0.8585 & 0.9278 & 0.9298 \\
0.3$^{\star}$ & \textbf{0.8585} & \textbf{0.9280} & \textbf{0.9298} \\
0.5           & 0.8575 & 0.9281 & 0.9293 \\
0.7           & 0.8536 & 0.9280 & 0.9289 \\
\midrule
\multicolumn{4}{l}{\small $\Delta_{\max}$ across $\beta$: AUC $\pm$0.0006, AUPR $\pm$0.0005, BalAcc $\pm$0.0025} \\
\bottomrule
\end{tabular}
\vspace{6pt}
\begin{tabular}{rrrr}
\toprule
$\lambda$ & \textbf{BalAcc} & \textbf{AUC} & \textbf{AUPR} \\
\midrule
0.05           & 0.8296 & 0.9135 & 0.9200 \\
0.10           & 0.8319 & 0.9146 & 0.9208 \\
0.20$^{\star}$ & \textbf{0.8337} & \textbf{0.9163} & \textbf{0.9232} \\
0.40           & 0.8313 & 0.9152 & 0.9217 \\
0.60           & 0.8350 & 0.9166 & 0.9228 \\
\midrule
\multicolumn{4}{l}{\small $\Delta_{\max}$ across $\lambda$: AUC $\pm$0.0016, AUPR $\pm$0.0016, BalAcc $\pm$0.0027} \\
\bottomrule
\end{tabular}
\end{table}

The $\beta$ sweep demonstrates that ViPER's primary metrics are
largely insensitive to the gate weight across the range $[0.1, 0.7]$,
with AUC varying by no more than 0.0006.
This confirms that the residual gating mechanism provides a stable
correction signal rather than a brittle additive offset, and that the
choice of $\beta = 0.3$ is not a critically tuned value.
The $\lambda$ sweep reveals that performance peaks near $\lambda = 0.2$
and degrades modestly at both extremes, consistent with the expectation
that the packing head acts as an auxiliary regulariser: too small a
weight provides insufficient auxiliary gradient, while too large a
weight risks allowing the packing objective to interfere with the
primary malware classification head.
Taken together, these results confirm that the selected hyperparameter
values are well-positioned within a stable operating region, and that
the reported ViPER performance is not an artefact of precise
hyperparameter tuning.

%=============================================================================
\section{Conclusion and Future Work}
\label{sec:conclusion}
%=============================================================================

We presented ViPER, a packing-aware dual-head vision
transformer for visualization-based malware detection. By
treating packing detection as a learned auxiliary task and
coupling its output to the primary malware classification head
through a residual gating mechanism, ViPER addresses a
structural limitation that has persisted across the body of
prior work in this area: the uniform treatment of packed and
unpacked executables under a single shared decision boundary.
Parameter-efficient fine-tuning via LoRA adapts the
self-supervised ViT-B/14 backbone to the byteplot domain
while keeping the trainable parameter count to 1.69\% of
the full model. Evaluated on 200,000 Windows PE byteplot
images, ViPER achieves a ROC-AUC of 0.9260 and AUPR
of 0.9279, outperforming all evaluated convolutional and
transformer baselines across every reported metric while
training only 1.49M parameters. Controlled ablation validates
the contribution of each architectural component, and the
packing detection auxiliary head achieves an AUC of 0.9949,
confirming that byteplot visual features carry strong
discriminative signal for packing state inference.

Several directions remain open for future work.
First, ViPER currently frames malware detection as a binary classification
problem. A natural extension is to incorporate fine-grained malware family
classification as an additional supervised head, enabling simultaneous
detection, packing-state inference, and family attribution within a single
unified framework.
Second, identifying the specific packer or protector type rather than a binary
packed/unpacked label could further refine the gating signal, as distinct
packers produce visually distinguishable byteplot textures that carry
discriminative information beyond a scalar packing confidence.
Third, exploring adaptive row-width representations that preserve section
boundaries could enrich the visual features available to the backbone beyond
the single fixed row width used in the current implementation.

%=============================================================================
\section*{Acknowledgements}
%=============================================================================

The authors would like to thank the Department of Computer and Information
Sciences at PIEAS and the Bioinformatics Lab at PIEAS for providing access
to the computational resources used in this work.

\printcredits

%% Bibliography
\bibliographystyle{cas-model2-names}

\bio{fatima_photo}
Fatima Qaiser is an undergraduate student in the Department of Computer and Information Sciences at the Pakistan Institute of Engineering and Applied Sciences (PIEAS), Islamabad, Pakistan. Her research focus is on cybersecurity, malware analysis, and software systems. This includes visualization-based threat detection, generative AI and its applications in security, and explainable AI (XAI).
\endbio

\bio{bisma_photo}
Bisma Tahir is an undergraduate student in the Department of Computer and Information Sciences at the Pakistan Institute of Engineering and Applied Sciences (PIEAS), Islamabad, Pakistan. Her research interests include deep learning, cybersecurity, computer architecture, generative AI applications, and explainable AI.
\endbio

\vskip1pc

\bio{abid_photo}
Muhammad Abid received the Ph.D. degree in Computer Science from Tsinghua University, China, in 2012. He is currently a Principal Scientist/ Associate Professor with the Pakistan Institute of Engineering and Applied Sciences (PIEAS). He has organized number of hands-on workshops on Malware Analysis, Network Penetration Testing, and Intrusion Analysis. He established Nvidia GPU Education Centre, Data Science Lab, and DIL Lab in PIEAS. His current research interests include the data security and privacy, malware analysis, and network security.
\endbio

\vskip1pc

\bio{nauman_photo}

Dr. Nauman Shamim is a faculty member in the Department of 
Computer and Information Sciences at PIEAS, Islamabad, Pakistan. 
His research interests include cybersecurity, network security, 
and applied machine learning.
\endbio

\end{document}